\documentclass[11pt,a4paper]{article}

\usepackage[T1]{fontenc}
\usepackage{lmodern}
\usepackage[a4paper,margin=1in]{geometry}
\usepackage{amsmath,amssymb,amsthm,bm}
\usepackage{graphicx,booktabs,multirow}
\usepackage[plain,noend]{algorithm2e}
\usepackage[round,authoryear]{natbib}
\usepackage{microtype}
\usepackage[hidelinks]{hyperref}

\newtheorem{theorem}{Theorem}
\newtheorem{proposition}{Proposition}
\newtheorem{corollary}{Corollary}

\newtheorem{assumption}{Assumption}
\theoremstyle{remark}

\newenvironment{keywords}{\par\medskip\noindent\textbf{Keywords: }\ignorespaces}{\par\medskip}
\hypersetup{
  pdftitle={Variational Bayes for Variable Selection with Adaptive Sharing across Populations and Responses},
  pdfauthor={Yuehan Yang, Pengfei Chen, Fan Yang, Chong You}
}
\begin{document}

\begin{center}
{\LARGE\bfseries Variational Bayes for Variable Selection with
Adaptive Sharing across Populations and Responses\par}
\vspace{1.2em}
{\large Yuehan Yang$^{1,2}$ \quad Pengfei Chen$^{1,2}$ \quad
Fan Yang$^{3,*}$ \quad Chong You$^{1,2,*}$\par}
\vspace{0.7em}
{\small $^{1}$Center for Mathematics and Interdisciplinary Sciences,
Fudan University, Shanghai 200433, China\par}
{\small $^{2}$Shanghai Institute for Mathematics and Interdisciplinary Sciences
(SIMIS), Shanghai 200433, China\par}
{\small $^{3}$School of Public Health, Shandong University,
Jinan 250012, China\par}
\vspace{0.6em}
{\footnotesize\texttt{24114020017@m.fudan.edu.cn} \quad
\texttt{24114020003@m.fudan.edu.cn}\par}
{\footnotesize\texttt{fanyang@sdu.edu.cn} \quad
\texttt{chongy@simis.cn}\par}
\vspace{0.45em}
{\footnotesize $^{*}$Corresponding authors: Fan Yang
(\texttt{fanyang@sdu.edu.cn}) and Chong You
(\texttt{chongy@simis.cn}).\par}
\end{center}

\begin{abstract}

We propose a variational Bayesian framework for variable selection with multiple responses across heterogeneous populations. The prior dependence among the inclusion indicators allows variable-selection information to be shared at different levels across populations and responses, while regression coefficients remain population- and response-specific. When the appropriate sharing structure is unknown, a data-driven rule adaptively determines the scope of information sharing from marginal association evidence. The resulting variational algorithm is computationally scalable and can be implemented using either individual-level or summary data. Under regularity conditions, we establish selection consistency and convergence properties of the variational parameters at fixed iterations of the algorithm for fixed dimension case. In high dimensions, we establish model-selection
consistency and coefficient contraction in both $\ell_1$ and $\ell_2$ norms. Simulation studies show competitive variable selection performance and demonstrate the potential gains from joint analysis. We further illustrate the summary-data implementation using genome-wide association summary statistics from the 2021 Global Lipids Genetics Consortium study for four lipid traits across five populations.

\end{abstract}


\section{Introduction}
\label{sec:introduction}
Variable selection aims to identify the predictors associated with a response from a set of candidates. When several responses are observed in multiple populations, the relevant predictors may overlap across responses and populations. Such overlap provides an opportunity to use evidence from related regressions rather than analyze each one in isolation. This can be useful when many predictors are considered but individual samples provide limited information for selection. However, the relevant predictors need not be the same across responses or populations. The statistical problem is therefore to combine information while identifying both shared and population- or response-specific associations.

Multivariate regression addresses the response dimension of this problem. Penalized methods such as remMap encourage both coefficient-level sparsity and selection of predictors associated with several responses \citep{Peng2010}. Bayesian methods represent selection through latent variables and structured priors. \citet{Brown1998} developed multivariate Bayesian variable selection with predictor-level binary indicators. Subsequent hierarchical models allow group-level and within-group selection \citep{Liquet2017}, while \citet{Canida2025} introduced an additional response-specific selection structure. More recently, mvSuSiE extends the sum-of-single-effects approach \citep{Wang2020SuSiE} to learn patterns of effect sharing across traits \citep{Zou2026MvSuSiE}.

Related information can also be obtained from data collected in other populations. Transfer-learning methods use auxiliary samples to improve inference for a target population while allowing differences between regression models. Trans-Lasso controls source–target coefficient differences to guide information transfer \citep{LiCaiLi2022}. In a Bayesian formulation, CONCERT uses conditional spike-and-slab priors to identify covariate-specific similarities between source and target parameters \citep{ZhangEtAl2026CONCERT}. For polygenic prediction, PRS-CSx shares continuous-shrinkage scales across populations while retaining population-specific effects \citep{Ruan2022}. Methods combining both populations and responses have also been developed. For example, \citet{Zhou2023MGflashfm} developed MGflashfm for joint fine-mapping
across multiple traits and population groups, with a multi-group prior
that requires variant overlap between at least one pair of group-specific models.

When variable inclusion can differ across responses and populations, Bayesian analysis involves a large collection of possible inclusion patterns. Posterior inference must also account for the regression coefficients and residual covariance matrices. These computational demands motivate posterior approximation. Variational Bayes (VB) approximates the posterior by optimizing over a tractable family of distributions. \citet{Ormerod2017} developed a mean-field approach to variable selection in linear regression, and \citet{You2023} studied an approximated collapsed approach. For multiple responses, \citet{Ruffieux2017} developed variational inference for multivariate regression. The use of such approximations also calls for an analysis of their statistical properties. \citet{RaySzabo2022} established contraction and prediction guarantees for sparse high-dimensional regression, while \citet{ZhangGao2020} developed general convergence theory that accounts for variational approximation error. These studies provide a basis for examining both computation and statistical behavior in the joint regression problem considered here.

In this paper, we propose MVP-VB, a variational Bayesian method for sparse multivariate regression across multiple populations. Information is shared through prior probabilities for population- and response-specific inclusion indicators. The prior allows both overlapping and nonoverlapping active sets across populations. We derive an iterative variational algorithm and a corresponding implementation based on cross-product summary statistics. We also propose a data-driven rule for choosing among the available prior-sharing structures.
Theoretical results are developed for both fixed- and high-dimensional models. Under regularity conditions, the fixed-dimensional analysis establishes variable-selection and coefficient consistency and characterizes the large-sample behavior of the variational iterates. In high dimensions, we establish model-selection consistency and coefficient contraction in both $(\ell_1)$ and $(\ell_2)$ norms for the globally optimal variational posterior under regularity conditions.
Simulation studies evaluate variable selection under different scenarios and we further illustrate the summary-data implementation on the real data
from the Global Lipids Genetics Consortium.

\section{Model and Variational Inference}
\label{sec:method}

\subsection{Model and prior specifications}
\label{subsec:mvp-model}

For population $k=1,\ldots,K$, let
$\boldsymbol Y_k\in\mathbb R^{n_k\times q}$ contain the observations on
$q$ responses and let $\boldsymbol X_k\in\mathbb R^{n_k\times m}$ be the
design matrix. The predictor and response indices are aligned across
populations, while the sample sizes $n_k$ may differ. We consider
\begin{equation}
\begin{aligned}
\boldsymbol Y_k
  &=\boldsymbol X_k(\boldsymbol Z_k\odot\boldsymbol B_k)
    +\boldsymbol{\mathcal E}_k,\\
\operatorname{vec}(\boldsymbol{\mathcal E}_k)\mid\boldsymbol\Sigma_k
  &\sim\mathcal N_{n_kq}
     (\boldsymbol0,\boldsymbol\Sigma_k\otimes\boldsymbol I_{n_k}),
\end{aligned}
\label{eq:model}
\end{equation}
where $\boldsymbol B_k\in\mathbb R^{m\times q}$ and
$\boldsymbol Z_k\in\{0,1\}^{m\times q}$. Throughout, $\operatorname{vec}$
stacks matrix columns, and $\odot$ and $\otimes$ denote the Hadamard and
Kronecker products. The product $\boldsymbol Z_k\odot\boldsymbol B_k$ is
the effective coefficient matrix. In particular, $Z_{kir}=0$ excludes
predictor $i$ from the regression for response $r$ in population $k$.
Each population has its own residual covariance matrix
$\boldsymbol\Sigma_k\succ\boldsymbol0$.

To share variable-selection information across populations and responses,
we link inclusion indicators through shared prior probabilities.
We consider an independent baseline and two hierarchical specifications:
\begin{equation}
\begin{array}{lll}
\text{Option 1:}&
 Z_{kir}\sim\operatorname{Bernoulli}(\pi), & 0<\pi<1,\\[2mm]
\text{Option 2:}&
 Z_{kir}\mid\pi_{ir}\sim\operatorname{Bernoulli}(\pi_{ir}),&
 \pi_{ir}\sim\operatorname{Beta}(a,b),\\[2mm]
\text{Option 3:}&
 Z_{kir}\mid\pi_i\sim\operatorname{Bernoulli}(\pi_i),&
 \pi_i\sim\operatorname{Beta}(a,b),
\end{array}
\label{eq:mvp-inclusion-priors}
\end{equation}
where $a,b>0$. The indicators are independent under Option~1 and
conditionally independent given the shared probabilities under
Options~2 and~3. Under Option~2, the common probability $\pi_{ir}$ links inclusion
indicators across populations for predictor $i$ and response $r$.
Integrating out $\pi_{ir}$ gives
\begin{equation*}
\Pr\!\left\{
Z_{kir}=1\mid(Z_{\ell ir})_{\ell\ne k}
\right\}
=
\frac{a+\sum_{\ell\ne k}Z_{\ell ir}}
     {a+b+K-1}.
\end{equation*}
This conditional probability increases with the number of other
populations in which the same predictor--response pair is included.
Option~3 extends this construction by sharing
$\pi_i$ among all $Kq$ inclusion indicators for predictor $i$.

For each population, we specify the following priors for the coefficient matrix and residual covariance matrix:
\begin{equation}
\begin{aligned}
\operatorname{vec}(\boldsymbol B_k)\mid\boldsymbol\Sigma_k
&\sim\mathcal N_{mq}
  (\boldsymbol0,\boldsymbol\Sigma_k\otimes\sigma_B^2\boldsymbol I_m),\\
\boldsymbol\Sigma_k
&\sim\mathcal{IW}_q(\nu,\boldsymbol\Psi),
\end{aligned}
\label{eq:mvp-slab-prior}
\end{equation}
with $\sigma_B^2>0$, $\nu>q-1$, and $\boldsymbol\Psi\succ\boldsymbol0$. Information sharing is therefore primarily introduced through the inclusion structure. 

Given the inclusion indicators and hyperparameters, coefficient inference for population $k$ uses its own data and residual covariance. The matrix-normal prior retains dependence among coefficients for different responses within that population.

\subsection{Variational inference}
\label{subsec:mvp-variational}
Posterior inference under the model in Section~\ref{subsec:mvp-model}
requires summation over the inclusion configurations together with integration over the continuous model parameters. These calculations become increasingly demanding as the number of candidate predictors increases. We therefore use variational Bayes (VB) to approximate the exact posterior $p(\boldsymbol\Theta\mid\boldsymbol Y,\boldsymbol X)$
by a tractable variational distribution $q(\boldsymbol\Theta)$. Here $\boldsymbol\Theta$ collects all unknown parameters and latent variables in the model.

The approximation is fitted by maximizing the evidence lower bound (ELBO)
\begin{equation}
\mathcal L(q)
=\mathbb E_q\!\left[\log p(\boldsymbol Y,\boldsymbol\Theta\mid\boldsymbol X)\right]
-\mathbb E_q\!\left[\log q(\boldsymbol\Theta)\right].
\label{eq:mvp-elbo}
\end{equation}
Maximizing \eqref{eq:mvp-elbo} is equivalent to minimizing the Kullback--Leibler divergence from 
$q(\boldsymbol\Theta)$ to the exact posterior
$p(\boldsymbol\Theta\mid\boldsymbol Y,\boldsymbol X)$.
Under the mean-field approximation \citep{Blei2017},
$q(\boldsymbol\Theta)=\prod_j q_j(\boldsymbol\theta_j)$,
the optimal variational factors $q_j^*$ satisfy
\begin{equation*}
q_j^\ast(\boldsymbol\theta_j)
\propto
\exp\left\{
\mathbb E_{q_{-j}^\ast}
\left[
\log p(\boldsymbol Y,\boldsymbol\Theta\mid\boldsymbol X)
\right]
\right\},
\end{equation*}
where $\boldsymbol\theta_j$ denotes a parameter block and
$q_{-j}^\ast=\prod_{\ell\ne j}q_\ell^\ast$.

Under Option~1, we use the mean-field factorization
\begin{equation*}
q(\boldsymbol\Theta)
=
\prod_{k=1}^K
\left\{
q(\boldsymbol b_k)q(\boldsymbol\Sigma_k)
\prod_{i=1}^m\prod_{r=1}^q q(Z_{kir})
\right\},
\end{equation*}
where $\boldsymbol b_k=\operatorname{vec}(\boldsymbol B_k)$.
Under Options~2 and~3, this product additionally includes
$\prod_{i=1}^m\prod_{r=1}^q q(\pi_{ir})$ and
$\prod_{i=1}^m q(\pi_i)$, respectively. 
Write
$\boldsymbol y_k=\operatorname{vec}(\boldsymbol Y_k)$,
$\boldsymbol z_k=\operatorname{vec}(\boldsymbol Z_k)$,
$W_{kir}=q(Z_{kir}=1)$,
and $\boldsymbol W_k=(W_{kir})$.
Define $\boldsymbol\Psi_k = \mathbb{E}_{q}(\boldsymbol\Sigma_k^{-1})$ and  \begin{equation}\label{eq:mvp-moments}
\begin{aligned}
\boldsymbol w_k  &= \operatorname{vec}(\boldsymbol W_k), \qquad
\boldsymbol w_k^M = \operatorname{diag}(\boldsymbol w_k), \\
\boldsymbol\Omega_k &= \mathbb{E}_{q^\ast}(\boldsymbol z_k\boldsymbol z_k^\top)
= \boldsymbol w_k\boldsymbol w_k^\top
+ \operatorname{diag}\left\{\boldsymbol w_k\odot(\boldsymbol{1}-\boldsymbol w_k)\right\}.
\end{aligned}
\end{equation}
The optimal variational factor for $\boldsymbol b_k$ is $q^\ast(\boldsymbol b_k)
=\mathcal N_{mq}(\boldsymbol m_k,\boldsymbol V_k)$, where
\begin{equation}
\begin{aligned}
\boldsymbol V_k^{-1}
&=
\boldsymbol\Psi_k\otimes\sigma_B^{-2}\boldsymbol I_m
+
\left\{
\boldsymbol\Psi_k\otimes
(\boldsymbol X_k^\top\boldsymbol X_k)
\right\}
\odot\boldsymbol\Omega_k,\\
\boldsymbol m_k
&=
\boldsymbol V_k
\boldsymbol w_k^M
(\boldsymbol\Psi_k\otimes\boldsymbol X_k^\top)
\boldsymbol y_k.
\end{aligned}
\label{eq:mvp-gaussian-update}
\end{equation}
The optimal variational factor for $\boldsymbol\Sigma_k$ is
$q^\ast(\boldsymbol\Sigma_k)
=\mathcal{IW}_q
(\nu+n_k+m,\boldsymbol T_k)$, where
\begin{equation}
\begin{aligned}
\boldsymbol T_k
={}&
\boldsymbol\Psi
+
\mathbb E_{q^\ast}
\left[
\{
\boldsymbol Y_k
-\boldsymbol X_k
(\boldsymbol Z_k\odot\boldsymbol B_k)
\}^{\top}
\{
\boldsymbol Y_k
-\boldsymbol X_k
(\boldsymbol Z_k\odot\boldsymbol B_k)
\}
\right]+
\sigma_B^{-2}
\mathbb E_{q^\ast}
(\boldsymbol B_k^\top\boldsymbol B_k),
\\
\boldsymbol\Psi_k
={}&
(\nu+n_k+m)\boldsymbol T_k^{-1}.
\end{aligned}
\label{eq:mvp-covariance-update}
\end{equation}
The optimal variational factor for each inclusion indicator is
\begin{equation*}
q^\ast(Z_{kir})
=
\operatorname{Bernoulli}(W_{kir}),
\quad W_{kir}=\operatorname{expit}(\lambda_{kir}+\delta_{kir}),
\end{equation*}
where
\begin{equation}
\begin{aligned}
\delta_{kir}
={}&
-\frac12(\boldsymbol\Psi_k)_{rr}
 \|\boldsymbol X_{ki}\|_2^2
 \left\{M_{kir}^2+(\boldsymbol V_{k(r,r)})_{ii}\right\}
+M_{kir}\boldsymbol X_{ki}^\top
 \boldsymbol Y_k(\boldsymbol\Psi_k)_{\cdot r}
\\
&-(\boldsymbol\Psi_k)_{rr}
 \boldsymbol X_{ki}^\top\boldsymbol X_{k,-i}
 \left[
  \left\{M_{kir}\boldsymbol M_{k,-i,r}
        +(\boldsymbol V_{k(r,r)})_{-i,i}\right\}
  \odot\boldsymbol W_{k,-i,r}
 \right]
\\
&-\boldsymbol X_{ki}^\top\boldsymbol X_k
 \left[
  \left\{M_{kir}\boldsymbol M_{k,-r}
       +\left((\boldsymbol V_{k(r,s)})_{i,\cdot}^{\top}\right)_{s\ne r}
  \right\}\odot\boldsymbol W_{k,-r}
 \right](\boldsymbol\Psi_k)_{-r,r}
\end{aligned}
\label{eq:mvp-delta-expanded}
\end{equation}
and 
\begin{equation}
\begin{aligned}
 \lambda_{kir}=&
\begin{cases}
\displaystyle
\log\frac{\pi}{1-\pi},
&\text{Option 1},\\[3mm]
\displaystyle
\psi\left(a+\sum_{\ell=1}^K W_{\ell ir}\right)
-\psi\left(b+K-\sum_{\ell=1}^K W_{\ell ir}\right),
&\text{Option 2},\\[4mm]
\displaystyle
\psi\left(a+\sum_{s=1}^q\sum_{\ell=1}^K W_{\ell is}\right)
-\psi\left(b+Kq-\sum_{s=1}^q\sum_{\ell=1}^K W_{\ell is}\right),
&\text{Option 3}.
\end{cases}
\label{eq:mvp-prior-logodds}
\end{aligned}
\end{equation}
Here $\psi$ denotes the digamma function, $\boldsymbol X_{ki}$ is column $i$ of $\boldsymbol X_k$, and
$\boldsymbol V_{k(r,s)}$ is the $m\times m$ covariance block for
responses $r$ and $s$. The subscripts $-i$ and $-r$ indicate deletion
of predictor $i$ and response $r$, respectively.
The matrix $((\boldsymbol V_{k(r,s)})_{i,\cdot}^{\top})_{s\ne r}$
has one column for each response $s\ne r$. The matrix $\boldsymbol M_k\in\mathbb R^{m\times q}$ is defined by $\operatorname{vec}(\boldsymbol M_k)=\boldsymbol m_k$.
Detailed derivations are given in
the Supplementary Material.

\subsection{Data-driven selection of the sharing structure}
\label{subsec:mvp-automatic}

When the sharing structure (\ref{eq:mvp-inclusion-priors}) is not
specified in advance, we propose an automatic mode (Option~4) which
uses marginal association evidence to choose among Options~1--3.
Candidate support for Option~2 is identified when the same
predictor--response association has sufficiently consistent evidence
across populations, subject to local-support and opposite-direction
safeguards. Candidate support for Option~3 is identified when the same
predictor has strong evidence for another response together with
supporting evidence for the current response. These candidate
associations are used to construct cross-population and
cross-response sharing groups, with Option~3 taking priority over
Option~2 when the two structures overlap. A global evidence rule then
determines whether the shared groups are retained; otherwise all
associations use Option~1. The resulting sharing structure is fixed
before the variational iterations. 

Let $o_{\ell ir}\in\{1,2,3\}$ denote the resulting fixed option assignment
for predictor $i$, response $r$, and population $\ell$. During the indicator updates at iteration $t$, $\boldsymbol W^{\mathrm{temp}}$ denotes the current working
inclusion probabilities, initialized at $\boldsymbol W^{(t)}$. For an Option~2
group associated with $(i,r)$, define 
\[
J_{ir,\mathrm{opt2}}
=
\#\{\ell:o_{\ell ir}=2\},
\qquad
u_{ir,\mathrm{opt2}}^{\mathrm{temp}}
=
\sum_{\ell:o_{\ell ir}=2}
W_{\ell ir}^{\mathrm{temp}}.
\]
For an Option~3 group associated with predictor $i$, define
\[
J_{i,\mathrm{opt3}}
=
\#\{(\ell,r):o_{\ell ir}=3\},
\qquad
u_{i,\mathrm{opt3}}^{\mathrm{temp}}
=
\sum_{(\ell,r):o_{\ell ir}=3}
W_{\ell ir}^{\mathrm{temp}}. 
\]
The prior log-odds under Option~4 are then
\begin{equation}
\lambda_{kir}^{\mathrm{temp}}
=
\begin{cases}
\displaystyle
\log\frac{\pi}{1-\pi},
& o_{kir}=1,\\[3mm]
\displaystyle
\psi\!\left(a+u_{ir,\mathrm{opt2}}^{\mathrm{temp}}\right)
-
\psi\!\left(
b+J_{ir,\mathrm{opt2}}
-u_{ir,\mathrm{opt2}}^{\mathrm{temp}}
\right),
& o_{kir}=2,\\[3mm]
\displaystyle
\psi\!\left(a+u_{i,\mathrm{opt3}}^{\mathrm{temp}}\right)
-
\psi\!\left(
b+J_{i,\mathrm{opt3}}
-u_{i,\mathrm{opt3}}^{\mathrm{temp}}
\right),
& o_{kir}=3.
\end{cases}
\label{eq:mvp-option4-logodds}
\end{equation}
Thus, the updates depend only on the members of each
selected sharing group. The full-group expressions in
\eqref{eq:mvp-prior-logodds} are recovered when
$J_{ir,\mathrm{opt2}}=K$ or $J_{i,\mathrm{opt3}}=Kq$,
respectively. When population sample sizes are highly unequal, marginal
association evidence is rescaled for the sharing decisions
without altering the likelihood inputs, as detailed in the
Supplementary Material.

Option~4 does not define an additional prior specification; rather, it
selects the sharing structure from the three priors in
\eqref{eq:mvp-inclusion-priors}. Unless otherwise stated, the full
MVP-VB analyses in both the simulation studies and the
multi-population real-data application use Option~4.

Conditional on a specified sharing structure, we use coordinate-ascent variational inference to optimize the variational distributions. The Gaussian and residual-covariance updates are the same for all four options. The prior log-odds are given by \eqref{eq:mvp-prior-logodds} under Options~1--3 and by
\eqref{eq:mvp-option4-logodds} under Option~4. Algorithm~\ref{alg:mvp-cavi} gives the update order, with $(t)$ denoting the iteration. After each complete update, the ELBO is evaluated at the current variational iterate, see the Supplementary Material for details. Iteration stops when the absolute difference between successive ELBO values is less than a prescribed tolerance $\epsilon$.

\begin{algorithm}[!htbp]
\caption{Coordinate-ascent variational updates for MVP-VB}
\label{alg:mvp-cavi}

\KwIn{
$\{\boldsymbol X_k,\boldsymbol Y_k\}_{k=1}^K$;
prior specification among Options~1--3,
or fixed Option~4 assignments and sharing groups;
hyperparameters; tolerance $\epsilon$
}

Set $t=1$ and initialize
$\boldsymbol W_k^{(1)}\in[0,1]^{m\times q}$ and
$\boldsymbol\Psi_k^{(0)}=\nu\boldsymbol\Psi^{-1}$
for $k=1,\ldots,K$\;

\Repeat{
two successive ELBO values differ by less than $\epsilon$
}{

\For{$k=1,\ldots,K$}{

Form $\boldsymbol w_k^{(t)}$ and
$\boldsymbol\Omega_k^{(t)}$
using \eqref{eq:mvp-moments}\;

Update $(\boldsymbol m_k^{(t)},\boldsymbol V_k^{(t)})$
by \eqref{eq:mvp-gaussian-update}, using
$\boldsymbol\Psi_k^{(t-1)}$ and $\boldsymbol W_k^{(t)}$\;

Update $q_{\Sigma_k}^{(t)}$ and
$\boldsymbol\Psi_k^{(t)}$
by \eqref{eq:mvp-covariance-update}\;

}

Set $\boldsymbol W^{\mathrm{temp}}=\boldsymbol W^{(t)}$\;

\For{each $(k,i,r)$, in the order $k$, then $i$, then $r$}{

For coordinates assigned to Option~2 or~3, update the
corresponding Beta factor using its fixed sharing group\;

Compute $\lambda_{kir}$ from
\eqref{eq:mvp-prior-logodds} for Options~1--3
or \eqref{eq:mvp-option4-logodds} for Option~4,
and $\delta_{kir}$ from
\eqref{eq:mvp-delta-expanded}\;

Set
$W_{kir}^{\mathrm{temp}}
=\operatorname{expit}(\lambda_{kir}+\delta_{kir})$\;

}

Set
$\boldsymbol W^{(t+1)}=\boldsymbol W^{\mathrm{temp}}$
and
$\boldsymbol w_k^{(t+1)}
=\operatorname{vec}(\boldsymbol W_k^{(t+1)})$\;

Update all Beta factors associated with Option~2 or~3
sharing groups using $\boldsymbol W^{(t+1)}$\;

Evaluate $\mathcal L^{(t)}$ at the current variational iterate
using the Supplementary Material\;

$t\leftarrow t+1$\;

}

\KwOut{
The final variational factors, inclusion probabilities,
and effective coefficient moments
}

\end{algorithm}

\subsection{Inference from summary statistics}
\label{subsec:mvp-summary}

The updates depend on individual observations through their cross-product
matrices. If these are available from the same sample, write
\begin{equation}
\boldsymbol D_k=\frac{\boldsymbol X_k^\top\boldsymbol X_k}{n_k},\qquad
\hat{\boldsymbol B}_k=\frac{\boldsymbol X_k^\top\boldsymbol Y_k}{n_k},\qquad
\boldsymbol L_k=\frac{\boldsymbol Y_k^\top\boldsymbol Y_k}{n_k}.
\label{eq:mvp-summary-inputs}
\end{equation}
Here $\hat{\boldsymbol B}_k$ denotes a cross-product summary, not the
joint regression coefficient estimate. The same algorithm applies after
replacing $\boldsymbol X_k^\top\boldsymbol X_k$,
$\boldsymbol X_k^\top\boldsymbol Y_k$, and
$\boldsymbol Y_k^\top\boldsymbol Y_k$ by
$n_k\boldsymbol D_k$, $n_k\hat{\boldsymbol B}_k$, and
$n_k\boldsymbol L_k$, respectively. The Supplementary Material gives the resulting
expressions.

For standardized $\boldsymbol X_k$ and $\boldsymbol Y_k$, $\boldsymbol D_k$ and $\boldsymbol L_k$ are correlation matrices and $\hat{\boldsymbol B}_k$ contains the univariate regression slopes.
In genetic applications, $\boldsymbol D_k$ is the population-specific
linkage disequilibrium matrix. External reference matrices may be used
when the corresponding sample cross-products are unavailable. The result
is then an approximate summary-data implementation rather than an individual-level analysis.

Summaries calculated from one sample satisfy
\begin{equation}
\begin{pmatrix}
\boldsymbol D_k&\hat{\boldsymbol B}_k\\
\hat{\boldsymbol B}_k^\top&\boldsymbol L_k
\end{pmatrix}
=\frac1{n_k}
\begin{pmatrix}\boldsymbol X_k&\boldsymbol Y_k\end{pmatrix}^{\!\top}
\begin{pmatrix}\boldsymbol X_k&\boldsymbol Y_k\end{pmatrix}
\succeq\boldsymbol0.
\label{eq:mvp-summary-compatibility}
\end{equation}
Their joint matrix has rank at most $n_k$. For external summaries, we
require the block positive-semidefiniteness condition in
\eqref{eq:mvp-summary-compatibility}.

\subsection{Implementation and variable selection}
\label{subsec:mvp-implementation}

The variational inclusion probability for a predictor--response pair is $W_{kir}$. For a specified threshold $c\in(0,1)$, define
\begin{equation*}
\hat\gamma_k(c)=\{(i,r):W_{kir}> c\}.
\end{equation*}
Throughout the MVP-VB algorithm, we set \(c = 0.5\).
The remaining numerical thresholds used in the simulation study and application are
reported in the Supplementary Material.
The coefficient estimate is the mean of the effective coefficient,
\begin{equation}
\mathbb E_q(Z_{kir}B_{kir})=W_{kir}M_{kir},
\label{eq:mvp-effective-mean}
\end{equation}
rather than $M_{kir}$ alone. Its variational variance is
\begin{equation}
\operatorname{Var}_q(Z_{kir}B_{kir})
 =W_{kir}(\boldsymbol V_{k(r,r)})_{ii}
  +W_{kir}(1-W_{kir})M_{kir}^2,
\label{eq:mvp-effective-variance}
\end{equation}
where $\boldsymbol V_{k(r,s)}$ denotes the $m\times m$ covariance block
between the $r$th and $s$th columns of $\boldsymbol B_k$. Derivations of
\eqref{eq:mvp-effective-mean}--\eqref{eq:mvp-effective-variance} are given in the Supplementary Material.

Initialization and computational screening used in the numerical studies are described in
the Supplementary Material as well.

\section{Theoretical Properties}
\label{sec:theory}

Let $\boldsymbol B_{k,0}\in\mathbb R^{m\times q}$ denote the true
effective coefficient matrix and let
$\boldsymbol\Sigma_{k,0}\in\mathbb R^{q\times q}$ denote the true
residual covariance matrix. The data-generating model is
\begin{equation}
\begin{aligned}
\boldsymbol Y_k&=\boldsymbol X_k\boldsymbol B_{k,0}
                    +\boldsymbol{\mathcal E}_k,\\
\operatorname{vec}(\boldsymbol{\mathcal E}_k)\mid\boldsymbol X_k
  &\sim N_{n_kq}(\boldsymbol0,
                  \boldsymbol\Sigma_{k,0}\otimes\boldsymbol I_{n_k}),
\end{aligned}
\label{eq:theory-sampling}
\end{equation}
where $\boldsymbol\Sigma_{k,0}\succ\boldsymbol0$.
For observation $\ell=1,\ldots,n_k$, let
$\boldsymbol x_{k\ell}\in\mathbb R^m$ and
$\boldsymbol\varepsilon_{k\ell}\in\mathbb R^q$ be the transposes of
row $\ell$ of $\boldsymbol X_k$ and $\boldsymbol{\mathcal E}_k$,
respectively. Within each population, the pairs
$(\boldsymbol x_{k\ell},\boldsymbol\varepsilon_{k\ell})$ are independent
and identically distributed, and the covariates are independent of the
errors. Write
$\boldsymbol G_k=\mathbb E(\boldsymbol x_{k1}\boldsymbol x_{k1}^{\top})$.
This is the population second-moment matrix, equal to the covariance
matrix when the covariates have mean zero. We write
$G_{k,ii}=(\boldsymbol G_k)_{ii}$.

The true inclusion matrix and coefficient vector are defined by
\begin{equation*}
Z_{kir,0}=\mathbf1\{B_{kir,0}\ne0\},\qquad
\boldsymbol Z_{k,0}=(Z_{kir,0}),\qquad
\boldsymbol b_{k,0}=\operatorname{vec}(\boldsymbol B_{k,0}).
\end{equation*}
The full and response-specific true active sets are
\begin{equation*}
\gamma_{k,0}=\{(i,r):B_{kir,0}\ne0\},\qquad
\gamma_{k,r,0}=\{i:(i,r)\in\gamma_{k,0}\}.
\end{equation*}
For a binary matrix $\boldsymbol Z\in\{0,1\}^{m\times q}$, define
\begin{equation*}
|\boldsymbol Z|=\sum_{i=1}^m\sum_{r=1}^q Z_{ir},\qquad
\boldsymbol\Gamma(\boldsymbol Z)
 =\operatorname{diag}\{\operatorname{vec}(\boldsymbol Z)\},\qquad
\boldsymbol\Gamma_k=\boldsymbol\Gamma(\boldsymbol Z_k).
\end{equation*}
Thus $|\boldsymbol Z_{k,0}|=|\gamma_{k,0}|$ counts active coefficient
coordinates, and $\boldsymbol\Gamma_k\boldsymbol b_k$ is the random
effective coefficient vector. Its true value is
$\boldsymbol b_{k,0}$, with
$\boldsymbol\Gamma(\boldsymbol Z_{k,0})\boldsymbol b_{k,0}
=\boldsymbol b_{k,0}$.
The subscript $0$ denotes a true value.

For a vector $\boldsymbol v$, $\boldsymbol v_S$ denotes the subvector
indexed by $S$. For a matrix $\boldsymbol A$, $\boldsymbol A_{S,T}$
denotes the submatrix with row set $S$ and column set $T$.
Coefficient coordinates follow the column-stacking convention:
$(i,r)$ corresponds to position $i+(r-1)m$.
The complement $\gamma_{k,0}^c$ is taken in
$\{1,\ldots,m\}\times\{1,\ldots,q\}$, whereas
$-\gamma_{k,r,0}$ is the complement of $\gamma_{k,r,0}$ in
$\{1,\ldots,m\}$.
When $S$ is a predictor set, $\boldsymbol X_{k,S}$ retains the columns
in $S$. Blocks indexed by response-specific active sets refer to the
corresponding coefficient coordinates.

For vectors, $\|\cdot\|_1$, $\|\cdot\|_2$, and $\|\cdot\|_\infty$
denote the $\ell_1$, Euclidean, and maximum norms, and
$\|\cdot\|_0$ counts nonzero coordinates.
For matrices, $\|\cdot\|_2$ denotes the
spectral norm. For symmetric matrices,
$\boldsymbol A\preceq\boldsymbol B$ means that
$\boldsymbol B-\boldsymbol A$ is positive semidefinite.
We use $u\vee v=\max(u,v)$, $u\wedge v=\min(u,v)$, and
$\lfloor u\rfloor$ for the integer part of $u$.
The relation $a_n\asymp b_n$ means that $a_n/b_n$ is bounded above
and away from zero by fixed constants for all sufficiently large $n$.

Throughout this section, $K$ and $q$ are fixed,
$n=\min_k n_k\to\infty$, and $n_k\asymp n$ for every $k$.
The inverse-Wishart hyperparameters are fixed, with $\nu>q-1$ and
$\boldsymbol\Psi\succ\boldsymbol0$.
Sampling probabilities and expectations under the true data-generating
parameters are denoted by $\Pr_{\theta_0}$ and
$\mathbb E_{\theta_0}$; they include randomness in the covariates
and errors. The notation $O_p$ and $o_p$ is with respect to this
sampling law. Vector and matrix stochastic orders in the main text
use the Euclidean and spectral norms, respectively.

\subsection{Fixed-dimensional behavior}
\label{subsec:fixed-theory}

The fixed-dimensional analysis concerns the iterates generated by
Algorithm~\ref{alg:mvp-cavi} under a specified choice among Options~1--4.
For each fixed $t\ge1$, we study their properties as the sample sizes
increase. The results do not require convergence of the algorithm to a
global variational optimum. Iteration notation follows
Section~\ref{subsec:mvp-variational}.

\begin{assumption}
For every population $k$:
\begin{enumerate}
\item[(A1)] The predictor dimension $m$ is fixed. The parameters
$\boldsymbol B_{k,0}$ and $\boldsymbol\Sigma_{k,0}$ do not vary with $n$,
$\boldsymbol G_k\succ\boldsymbol0$, and
$\mathbb E\|\boldsymbol x_{k1}\|_2^4<\infty$.
The Gaussian-slab variance is fixed, with $0<\sigma_B^2<\infty$.
\item[(A2)] Under Option~1, the common prior inclusion probability
$\pi=\pi_n$ satisfies $\pi_n\to0$, $n_k\pi_n\to0$,
$\frac{\log(1/\pi_n)}{n_k}\to 0$. Under Options~2 and~3, $a>0$ is fixed and the common Beta parameter $b=b_n$ satisfies $b_n\to\infty$, $\frac{n_k}{b_n}\to0$,$\frac{\log b_n}{n_k}\to0$.
The sequences are common to the populations; the displayed conditions
hold for every $k$. Under Option~4, both the conditions for Option~1 and those for Options~2 and~3 are imposed.
\end{enumerate}
\end{assumption}
Condition~(A1) rules out population-level linear degeneracy and controls
the fluctuations of the sample cross-products. Condition~(A2) balances
exclusion of null coordinates against retention of nonzero effects.
The prior probability of inclusion decreases with sample size, but its
logarithm is of smaller order than the likelihood evidence for a fixed
nonzero coefficient. In the hierarchical specifications, the same role
is played by the increasing Beta parameter $b_n$.

When $\gamma_{k,0}\ne\varnothing$, define
\begin{equation}
\boldsymbol C_{k,0}
  =\bigl[\boldsymbol\Sigma_{k,0}^{-1}\otimes\boldsymbol G_k
     \bigr]_{\gamma_{k,0},\gamma_{k,0}}.
\label{eq:theory-active-information}
\end{equation}
This is the population information matrix restricted to the true active
coefficient coordinates, with the residual covariance held at its true value.

\begin{theorem}
\label{res:fixed-main-result-2}
Suppose that Assumption~1 holds and fix $t\geq1$. The coordinate
updates may use any of Options~1--4. For every population $k$, suppose
that there are working sets $\gamma_{k,r}$ such that $\gamma_{k,r,0}\subseteq\gamma_{k,r}$ for $r=1,\ldots,q$. Write $\mathcal A_k=\{(i,r):i\in\gamma_{k,r},\ r=1,\ldots,q\}$ and suppose that
\[
\max_{1\leq r\leq q}\left\|\boldsymbol1-\boldsymbol W_{k,\gamma_{k,r},r}^{(t)}\right\|_\infty=o_p(n_k^{-1}),
\qquad
\max_{1\leq r\leq q}\left\|\boldsymbol W_{k,-\gamma_{k,r},r}^{(t)}\right\|_\infty=o_p(n_k^{-1}).
\]
All true coordinates are retained for the updates under consideration.

The continuous-parameter updates at iteration $t$ satisfy
\[
\begin{aligned}
(\boldsymbol m_k^{(t)})_{\mathcal A_k}
&=(\boldsymbol b_{k,0})_{\mathcal A_k}
  +O_p(n_k^{-1/2}),\\
(\boldsymbol V_k^{(t)})_{\mathcal A_k,\mathcal A_k}
&=\frac{1}{n_k}
 \left(
 [\boldsymbol\Psi_k^{(t-1)}\otimes
   \boldsymbol G_k]_{\mathcal A_k,\mathcal A_k}
 \right)^{-1}
 +O_p(n_k^{-3/2}),\\
\boldsymbol\Psi_k^{(t)}
&=\boldsymbol\Sigma_{k,0}^{-1}
  +O_p(n_k^{-1/2}).
\end{aligned}
\]
After all indicator coordinates have been updated once, there is a
constant $c_k>0$ such that
\[
\max_{(i,r)\in\gamma_{k,0}}
\{1-W_{kir}^{(t+1)}\}
=O_p\{\exp(-c_kn_k)\},
\qquad
\max_{1\leq r\leq q}\max_{i\notin\gamma_{k,r}}
W_{kir}^{(t+1)}
=O_p\{\exp(-c_kn_k)\},
\]
and
\[
n_k\max_{(i,r)\notin\gamma_{k,0}}
W_{kir}^{(t+1)}
\xrightarrow{p}0.
\]
Consequently, at the next continuous-parameter update,
\[
\begin{aligned}
\boldsymbol\Psi_k^{(t+1)}
&=\boldsymbol\Sigma_{k,0}^{-1}
  +O_p(n_k^{-1/2}),\\
(\boldsymbol w_k^{(t+1)}\odot
 \boldsymbol m_k^{(t+1)})_{\gamma_{k,0}}
&=(\boldsymbol b_{k,0})_{\gamma_{k,0}}
  +O_p(n_k^{-1/2}),\\
(\boldsymbol w_k^{(t+1)}\odot
 \boldsymbol m_k^{(t+1)})_{\gamma_{k,0}^{c}}
&=o_p(n_k^{-1}).
\end{aligned}
\]
When $\gamma_{k,0}$ is nonempty, its Gaussian covariance parameter
satisfies
\begin{equation}
\begin{aligned}
(\boldsymbol V_k^{(t+1)})_{\gamma_{k,0},\gamma_{k,0}}
&=n_k^{-1}\boldsymbol C_{k,0}^{-1}
  +O_p(n_k^{-3/2}),\\
(\boldsymbol V_k^{(t+1)})_{\gamma_{k,0},\gamma_{k,0}^{c}}
&=O_p(n_k^{-1}),\\
(\boldsymbol V_k^{(t+1)})_{\gamma_{k,0}^{c},\gamma_{k,0}^{c}}
&=O_p(1).
\end{aligned}
\label{eq:fd-gaussian-covariance}
\end{equation}
Expressions involving empty index sets are omitted, and a maximum over an empty set is defined as zero.
\end{theorem}

The full-model initialization $W_{kir}^{(1)}=1$ satisfies the neighborhood condition with $\gamma_{k,r}=\{1,\ldots,m\}$. For screened initializations, the result applies when the stated neighborhood condition holds and all true coordinates remain available for updating.

Theorem~\ref{res:fixed-main-result-2} shows that the next indicator update separates active and inactive coordinates near a correct working model. At the following update, active coefficient estimates and residual precision have $O_p(n_k^{-1/2})$ errors, while inactive effective coefficient estimates are $o_p(n_k^{-1})$.

Note that the covariance parameter $\boldsymbol V_k^{(t+1)}$ in
\eqref{eq:fd-gaussian-covariance} describes the coefficient vector $\boldsymbol b_k$, whose variances need not vanish when their indicators are excluded. For the effective coefficient vector
$\boldsymbol\Gamma_k\boldsymbol b_k$, Theorem~\ref{res:fixed-main-result-2} and the covariance identity in the Supplementary Material give,
\begin{equation}
\begin{aligned}
\bigl\{\operatorname{Cov}_{q^{(t+1)}}(\boldsymbol\Gamma_k\boldsymbol b_k)
 \bigr\}_{\gamma_{k,0},\gamma_{k,0}}
  &=n_k^{-1}\boldsymbol C_{k,0}^{-1}+o_p(n_k^{-1}),\\
\bigl\{\operatorname{Cov}_{q^{(t+1)}}(\boldsymbol\Gamma_k\boldsymbol b_k)
 \bigr\}_{\gamma_{k,0}^{c},\gamma_{k,0}^{c}}
  &=o_p(n_k^{-1}).
\end{aligned}
\label{eq:fd-effective-covariance}
\end{equation}
Thus, effective coefficients concentrate at zero on the inactive coordinates. The derivation is given in the Supplementary Material.

\begin{corollary}
\label{cor:fd-threshold-selection}
Under the conditions of Theorem~\ref{res:fixed-main-result-2}, fix
$t\ge1$ and $c\in(0,1)$. Define
\[
\hat\gamma_k^{(t+1)}(c)=\{(i,r):W_{kir}^{(t+1)}>c\}.
\]
Then
\begin{equation*}
\Pr_{\theta_0}\left\{
\hat\gamma_k^{(t+1)}(c)=\gamma_{k,0}
\text{ for all }k=1,\ldots,K\right\}\rightarrow1.
\end{equation*}
\end{corollary}
The corollary means that any fixed threshold $c\in(0,1)$, including $c=1/2$, consistently recovers the full support pattern across populations and responses. The proof of Theorem~\ref{res:fixed-main-result-2} and its supporting results are provided in the Supplementary Material.

Under Option~2, when the relevant inclusion weights are near zero,
support from one other population increases the target's prior
log-odds by approximately $1/a+1/(b+K-1)$; under Option~3, the
same expression holds with $K$ replaced by $Kq$.
Thus, when $a=1$ and $b$ is large, either update multiplies the
inclusion odds by approximately $e$, holding $\delta_{kir}$ fixed.
For a true association with $W_{kir}$ near one, one additional
source of support instead reduces $1-W_{kir}$ by a factor of
approximately $e^{-1/(a+1)}$ under the same comparison.
These constant-factor changes may improve detection of partially
shared associations in finite samples.
Because the number of populations is fixed, their sample sizes
are of the same order, and sharing changes the prior log-odds by
$O(1)$, it does not alter the $O_p(n_k^{-1/2})$ coefficient
estimation rate or the $n_k^{-1}$ order of the leading covariance.

\subsection{Comparison of the leading variational covariance}
\label{subsec:theory-covariance}

The matrix in \eqref{eq:theory-active-information} also describes how
joint response analysis changes the leading coefficient covariance.
Let $[\boldsymbol C_{k,0}^{-1}]_{r,r}$ denote the diagonal block
corresponding to the active predictors for response $r$.
For response $r$ analyzed alone with its true active set, the corresponding
leading covariance is
$n_k^{-1}(\boldsymbol\Sigma_{k,0})_{rr}
 \boldsymbol G_{k,\gamma_{k,r,0},\gamma_{k,r,0}}^{-1}$.

\begin{proposition}
\label{prop:fixed-cov-dominance}
Under Assumption~(A1), for every
population $k$ and response $r$ with
$\gamma_{k,r,0}\ne\varnothing$,
\begin{equation*}
[\boldsymbol C_{k,0}^{-1}]_{r,r}
 \preceq
(\boldsymbol\Sigma_{k,0})_{rr}
 \boldsymbol G_{k,\gamma_{k,r,0},\gamma_{k,r,0}}^{-1}.
\end{equation*}
\end{proposition}
That means the joint response analysis would have no larger leading variational variance than separate analysis for any linear combination of a response's active coefficients. This is a classical result; see \citep{Zellner1962}. Any improvement depends on residual response correlation and the response-specific active sets. Equality holds, for example, when the residual covariance is diagonal or all responses have the same active set. The proof and general equality condition are given in the Supplementary Material.

\subsection{High-dimensional properties}
\label{subsec:high-dimensional-theory}

We now allow the predictor dimension $m=m_n$ to grow with the sample
sizes. The following results concern the individual-level model under a fixed
choice among Options~1--3 and the corresponding global mean-field
variational optimum. 

Let $\Pi_k(\,\cdot\mid\boldsymbol X,\boldsymbol Y)$ denote the
population-$k$ marginal of the exact posterior distribution of
$(\boldsymbol b_k,\boldsymbol\Sigma_k,\boldsymbol Z_k)$ under the
specified prior option. Define the mean-field family
\begin{equation}
\mathcal Q_k=\left\{q_k:
q_k(\boldsymbol b_k,\boldsymbol\Sigma_k,\boldsymbol Z_k)
 =q_k(\boldsymbol b_k)q_k(\boldsymbol\Sigma_k)
   \prod_{i=1}^m\prod_{r=1}^q q_k(Z_{kir})\right\}.
\label{eq:hd-mf-family}
\end{equation}

\begin{assumption}
For every population $k$, assume the following. Condition~(B1) is global whereas Conditions~(B2)--(B5) hold for every population $k$.
\begin{enumerate}
\item[(B1)] \textit{Existence of a variational optimum.}
Under Option~1, for every population $k$, the variational optimum
$q_k^\ast$ over $\mathcal Q_k$ exists. Under Options~2 and~3, a global
mean-field variational optimum $q^\ast$ under the factorization in
Section~\ref{subsec:mvp-variational} exists, and $q_k^\ast$ denotes its
marginal distribution of
$(\boldsymbol b_k,\boldsymbol\Sigma_k,\boldsymbol Z_k)$.

\item[(B2)] \textit{Sparsity and residual covariance.}
We have $mq>1$, $\log(mq)=o(n_k)$,
$1\le|\boldsymbol Z_{k,0}|=O(1)$, and
$\|\boldsymbol b_{k,0}\|_2^2=O(1)$.
For fixed $0<c_\Sigma<C_\Sigma<\infty$,
\[
c_\Sigma\boldsymbol I_q\preceq\boldsymbol\Sigma_{k,0}
 \preceq C_\Sigma\boldsymbol I_q.
\]

\item[(B3)] \textit{Prior sequences.}
The slab variance $\sigma_B^2$ may depend on $n$. There exists a
deterministic sequence $\delta_n\to\infty$, allowed to diverge
arbitrarily slowly, such that
\[
\delta_n\log(\sqrt{n_k}\vee mq)=o(n_k),\qquad
n_k\sigma_B^2\asymp n_k\vee(mq)^{2+3\delta_n}.
\]
Under Option~1, $\pi=\pi_n$ satisfies $\frac{\pi_n}{1-\pi_n}\asymp(mq)^{-1}$. Under Options~2 and~3, $a>0$ is fixed and the common Beta parameter
$b=b_n$ satisfies $b_n\asymp mq$.

\item[(B4)] \textit{Population sparse-design regularity.}
For a sufficiently large fixed $\nu_0>0$, define the cutoff
\[
m_{n_k}(\nu_0)=(mq)\wedge
 \frac{n_k}{(2+\nu_0)\log(mq)}.
\]
There are fixed $0<c_X<C_X<\infty$ and $K_X<\infty$ such that
$\max_iG_{k,ii}\le C_X$ and, for every nonzero
$\boldsymbol u\in\mathbb R^m$ with
$\|\boldsymbol u\|_0\le
 \min\{m,\lfloor m_{n_k}(\nu_0)\rfloor+1\}$,
\begin{equation*}
\boldsymbol u^\top\boldsymbol G_k\boldsymbol u
 \ge c_X\|\boldsymbol u\|_2^2,\qquad
\mathbb E\exp\left\{
\frac{(\boldsymbol u^\top\boldsymbol x_{k1})^2}
 {K_X^2\boldsymbol u^\top\boldsymbol G_k\boldsymbol u}
\right\}\le2.
\end{equation*}
\item[(B5)] \textit{Minimum signal strength.}
For a sufficiently large fixed $M_B>0$,
\begin{equation*}
\min_{(i,r)\in\gamma_{k,0}} B_{kir,0}^2
 \ge M_B\frac{|\boldsymbol Z_{k,0}|\,\delta_n
                  \log(\sqrt{n_k}\vee mq)}{n_k}.
\end{equation*}
\end{enumerate}
\end{assumption}

Condition~(B2) allows the candidate dimension to increase while keeping
the true support size bounded and the residual covariance well conditioned.
Condition~(B3) specifies the prior sequences; the restriction on
$\delta_n$ ensures that the contraction rates below vanish.
Condition~(B4) provides identifiability and tail control for sparse
covariate combinations while allowing the full sample Gram matrix to be singular.
Condition~(B5) permits nonzero effects to decrease with sample size, provided they remain above the noise and model-selection scale.

\begin{theorem}
\label{thm:hd-vb-model-selection}
Suppose that Assumption~2 holds. For every population $k$,
\begin{equation*}
\mathbb E_{\theta_0}q_k^\ast(\boldsymbol Z_k=\boldsymbol Z_{k,0})
 \rightarrow1.
\end{equation*}
Consequently,
$q_k^\ast(\boldsymbol Z_k=\boldsymbol Z_{k,0})\xrightarrow{p}1$ under
$\Pr_{\theta_0}$.
\end{theorem}

\begin{theorem}
\label{thm:hd-vb-contraction}
Suppose that Assumption~2 holds and define
\begin{equation*}
\varepsilon_{k,n_k}
 =\left\{\frac{\delta_n\log(\sqrt{n_k}\vee mq)}{n_k}\right\}^{1/2}.
\end{equation*}
Then $\varepsilon_{k,n_k}\to0$. For a sufficiently large fixed $M>0$,
\begin{equation*}
\begin{aligned}
\mathbb E_{\theta_0}q_k^\ast\left(
 \|\boldsymbol\Gamma(\boldsymbol Z_k)\boldsymbol b_k
        -\boldsymbol b_{k,0}\|_2
 >M\sqrt{|\boldsymbol Z_{k,0}|}\,\varepsilon_{k,n_k}
                         \right)&\rightarrow0,\\
\mathbb E_{\theta_0}q_k^\ast\left(
 \|\boldsymbol\Gamma(\boldsymbol Z_k)\boldsymbol b_k
        -\boldsymbol b_{k,0}\|_1
 >M|\boldsymbol Z_{k,0}|\,\varepsilon_{k,n_k}
                         \right)&\rightarrow0.
\end{aligned}
\end{equation*}
\end{theorem}
This result shows that the variational posterior for the effective coefficients concentrates around the true coefficient vector in both $\ell_1$ and $\ell_2$ norms. 
The proofs combine concentration of the exact posterior with control of the variational approximation error, following the approach of \citet{ZhangGao2020,RaySzabo2022}. Details are provided in the Supplementary Material.

Although the high-dimensional results are established for
fixed Options~1--3, Option~4 preserves their sparsity-penalty
scale. Under Condition~(B3), with fixed $K$ and $q$, its
prior log-odds satisfy
$\lambda_{kir}=-\log(mq)+O(1)$
uniformly over the selected sharing groups. This bounded
difference is negligible relative to the nonzero-signal
separation scale under Condition~(B5), while the coefficient
updates retain the same form. These properties suggest
similar high-dimensional selection and contraction behavior
under Option~4.

\section{Simulation Studies}
\label{sec:simulation}
\subsection{Simulation setting}
\label{subsec:simulation-design}

We evaluate the variable-selection performance of MVP-VB using
simulated genotypes and phenotypes. We also consider MVP-VB-pop,
which applies Option~1 separately to each population while modelling
all phenotypes jointly, and MVP-VB-pop-pheno, which fits each
population--phenotype combination separately. We further compare
MVP-VB with MGflashfm \citep{Zhou2023MGflashfm}, MultiSuSiE
\citep{Rossen2026MultiSuSiE}, mvSuSiE \citep{Zou2026MvSuSiE}, and
MSGLasso \citep{Li2015}. For MGflashfm and MultiSuSiE, the selected SNP--phenotype pairs
from their multi-group outputs are assigned to all populations
when computing population-specific $F_1$ scores.
 For MSGLasso, we use its no-group
multivariate lasso specification and refer to it as MVLasso. These
comparisons assess the contributions of cross-population and joint
phenotype analysis.

MVP-VB is implemented using Option~4 throughout. MGflashfm selects
SNP--phenotype pairs with multi-group marginal posterior probability
greater than \(0.90\), whereas MultiSuSiE analyzes each phenotype
jointly across populations and selects SNPs with overall posterior
inclusion probability greater than \(0.90\). For mvSuSiE, a
SNP--phenotype pair is selected when the SNP belongs to a retained
\(95\%\) credible set with minimum absolute correlation at least
\(0.5\) and the corresponding single effect has average
\(\mathrm{lfsr}<0.01\)
\citep{Zou2026MvSuSiE,Stephens2017}. MVLasso retains only the
elementwise \(\ell_1\) penalty, with all group penalties set to zero,
and its tuning parameter is selected by the extended Bayesian
information criterion \citep[EBIC,][]{ChenChen2008}. Implementation
and tuning details are provided in the Supplementary Material.

Genotype matrices \(\boldsymbol X_k\) are generated using HAPGEN2
\citep{Su2011HAPGEN2} with the 1000 Genomes Project Phase~3 reference
haplotypes \citep{1000Genomes2015} to preserve realistic linkage
disequilibrium (LD) patterns among SNPs. The same pre-generated
genotype matrices are used across replications for each combination
of \(K\) and \(m\). Genotypes are standardized within each population,
and the simulated phenotypes are centered. Causal SNPs are sampled
from distinct LD blocks within each population.

The residual correlation matrix has entries
\((\boldsymbol R_{\varepsilon})_{rs}=0.30^{|r-s|}\).
The absolute values of nonzero regression coefficients are drawn
independently from \(\operatorname{Unif}(0.60,1.00)\), with randomly
assigned signs. When the same SNP--phenotype pair is active in
multiple populations, its effect sign is shared among those
populations, while the effect magnitudes are generated separately.

The population-specific residual covariance matrix is
\(\boldsymbol\Sigma_{k,0}=\sigma_k^2\boldsymbol R_{\varepsilon}\),
where \(\sigma_k^2\) is chosen to attain the target multivariate
heritability. To achieve the target heritability, we first calculate the aggregate genetic variance \(V_{g,k} = \sum_{r=1}^{q} \hat{\operatorname{Var}}(\boldsymbol X_k \boldsymbol B_{k,0,\cdot r})\), with \(\hat{\operatorname{Var}}\) representing the sample variance across individuals and \(\boldsymbol B_{k,0,\cdot r}\) denoting the \(r\)-th column of the true coefficient matrix. We then set \(\sigma_k^2 = V_{g,k}(1-h^2)/(q h^2)\), which ensures that the population-specific heritability remains exactly \(h_k^2 = V_{g,k}/[V_{g,k}+\operatorname{tr}(\boldsymbol\Sigma_{k,0})] = h^2\).

We vary the number of populations \(K\) over \(\{2,4\}\) and the
number of SNPs \(m\) over \(\{500,1000\}\). The number of phenotypes
\(q\) varies over \(\{3,6\}\), and the target multivariate
heritability \(h^2\) over \(\{0.15,0.30\}\).
Let \(S\) denote the number of causal SNPs in each population.
For all three scenarios, we use \(S\in\{5,8\}\) when \(q=3\) and
\(S\in\{10,15\}\) when \(q=6\). Each causal SNP affects exactly one
phenotype within its population, so each population contains exactly
\(S\) nonzero SNP--phenotype pairs in every scenario.
The population sample sizes are \((200,400)\) for \(K=2\) and
\((200,300,400,500)\) for \(K=4\).

We consider the following three association structures:
\begin{description}

\item[Scenario~I.]
\textit{Population-specific associations.}
The causal SNP sets are mutually disjoint across populations.
Each causal SNP affects exactly one phenotype, and every phenotype
has at least one causal SNP in each population. The numbers of
causal SNPs assigned to different phenotypes within a population
differ by at most one.

\item[Scenario~II.]
\textit{Cross-population sharing.}
Each population contains \(S\) causal SNPs, each affecting exactly
one phenotype. The \(S\) causal signals are assigned to phenotypes 
such that every phenotype is represented, with the
numbers of assigned signals differing by at most one.

For \(K=2\), each causal signal is randomly assigned either a causal SNP
shared by both populations or two distinct population-specific
causal SNPs. The sharing pattern is sampled subject to the
constraint that every phenotype has at least one SNP--phenotype
pair shared by both populations.

For \(K=4\), each causal signal corresponds to a causal SNP shared by a randomly
selected subset of two, three, or four populations. Populations
outside this subset receive distinct population-specific causal
SNPs affecting the same phenotype.

\item[Scenario~III.]
\textit{Cross-population and cross-phenotype sharing.}
All \(S\) causal SNPs are shared by all \(K\) populations.
Within each population, each causal SNP affects exactly one
phenotype, while across populations it affects exactly two
distinct phenotypes.

For each causal SNP, two distinct phenotypes are randomly
selected. When \(K=2\), the two phenotypes are randomly assigned
to the two populations, one phenotype per population. When
\(K=4\), the populations are randomly divided into two groups
of two, with each group assigned to one of the two phenotypes.

The population-to-phenotype assignments are initially randomized
independently for each causal SNP, without using population
sample sizes or a fixed population ordering. The complete
assignment is resampled until every phenotype is represented
by at least one causal SNP in each population.

\end{description}

The simulation design comprises 96 distinct parameter
configurations across the three association structures.
For each configuration, 50 replications are generated using
the fixed genotype matrices, with independently generated
causal effects and residual noise across replications.

Variable-selection accuracy is evaluated at the
SNP--phenotype-pair level within each population using
\begin{equation*}
F_1=
\frac{2\,\mathrm{TP}}
{2\,\mathrm{TP}+\mathrm{FP}+\mathrm{FN}},
\end{equation*}
where \(\mathrm{TP}\), \(\mathrm{FP}\), and \(\mathrm{FN}\) are the
numbers of correctly selected nonzero coefficients, incorrectly
selected zero coefficients, and missed nonzero coefficients,
respectively. 
\subsection{Simulation results}
\label{subsec:simulation-results}

Table~\ref{tab:f1-m1000-q6-h30} reports population-specific
$F_1$ scores for $m=1000$, $q=6$, and $h^2=0.30$.
We report the mean $F_1$ scores over 50 replications, with Monte Carlo
standard errors (MCSEs) in parentheses. Boldface indicates the largest
mean $F_1$ among the compared methods for each population and
configuration; ties at the displayed precision are also highlighted.
For conciseness, the results for the remaining seven $(m,q,h^2)$ combinations are
provided in the Supplementary Material.

\begin{table*}[!t]
\centering
\caption{Mean $F_1$ (MCSE) for $m=1000$, $q=6$, and $h^2=0.30$.}
\label{tab:f1-m1000-q6-h30}
\footnotesize
\setlength{\tabcolsep}{1.8pt}
\renewcommand{\arraystretch}{1.05}
\begin{tabular}{llcccccc}
\toprule
& & \multicolumn{2}{c}{$K=2$}
    & \multicolumn{4}{c}{$K=4$} \\
\cmidrule(lr){3-4}\cmidrule(lr){5-8}
$S$ & Method
& Pop.~1 & Pop.~2
& Pop.~1 & Pop.~2 & Pop.~3 & Pop.~4 \\
\midrule
\multicolumn{8}{l}{\textit{I. Population-specific SNP--trait associations.}} \\
\midrule
\multirow{7}{*}{$10$} & MVP-VB & \textbf{0.946 (0.01)} & \textbf{0.938 (0.01)} & \textbf{0.934 (0.01)} & \textbf{0.934 (0.01)} & \textbf{0.939 (0.01)} & \textbf{0.941 (0.01)} \\
 & MVP-VB-pop & \textbf{0.946 (0.01)} & 0.937 (0.01) & 0.932 (0.01) & \textbf{0.934 (0.01)} & 0.937 (0.01) & 0.939 (0.01) \\
 & MVP-VB-pop-pheno & 0.856 (0.01) & 0.905 (0.01) & 0.852 (0.01) & 0.886 (0.01) & 0.909 (0.01) & 0.914 (0.01) \\
 & MGflashfm & 0.014 (0.01) & 0.014 (0.01) & 0.059 (0.01) & 0.070 (0.02) & 0.069 (0.01) & 0.090 (0.02) \\
 & MultiSuSiE & 0.626 (0.01) & 0.633 (0.01) & 0.374 (0.01) & 0.377 (0.01) & 0.387 (0.01) & 0.389 (0.01) \\
 & mvSuSiE & 0.902 (0.01) & 0.920 (0.01) & 0.913 (0.01) & 0.905 (0.01) & 0.925 (0.01) & 0.925 (0.01) \\
 & MVLasso & 0.003 (0.00) & 0.003 (0.00) & 0.003 (0.00) & 0.003 (0.00) & 0.003 (0.00) & 0.003 (0.00) \\
\addlinespace[3pt]
\multirow{7}{*}{$15$} & MVP-VB & 0.904 (0.01) & \textbf{0.946 (0.01)} & \textbf{0.903 (0.01)} & \textbf{0.917 (0.01)} & \textbf{0.920 (0.01)} & \textbf{0.930 (0.01)} \\
 & MVP-VB-pop & \textbf{0.905 (0.01)} & 0.944 (0.01) & \textbf{0.903 (0.01)} & 0.914 (0.01) & 0.919 (0.01) & 0.929 (0.01) \\
 & MVP-VB-pop-pheno & 0.848 (0.01) & 0.920 (0.01) & 0.842 (0.01) & 0.892 (0.01) & 0.896 (0.01) & 0.913 (0.01) \\
 & MGflashfm & 0.002 (0.00) & 0.004 (0.00) & 0.019 (0.01) & 0.025 (0.01) & 0.025 (0.01) & 0.041 (0.01) \\
 & MultiSuSiE & 0.520 (0.01) & 0.679 (0.01) & 0.262 (0.01) & 0.384 (0.01) & 0.396 (0.01) & 0.430 (0.01) \\
 & mvSuSiE & 0.707 (0.01) & 0.742 (0.01) & 0.725 (0.01) & 0.706 (0.01) & 0.723 (0.01) & 0.735 (0.01) \\
 & MVLasso & 0.005 (0.00) & 0.005 (0.00) & 0.005 (0.00) & 0.005 (0.00) & 0.005 (0.00) & 0.005 (0.00) \\
\midrule
\multicolumn{8}{l}{\textit{II. Partially shared SNP--trait associations.}} \\
\midrule
\multirow{7}{*}{$10$} & MVP-VB & \textbf{0.943 (0.01)} & \textbf{0.954 (0.01)} & \textbf{0.954 (0.01)} & \textbf{0.945 (0.01)} & \textbf{0.944 (0.01)} & \textbf{0.947 (0.01)} \\
 & MVP-VB-pop & 0.941 (0.01) & 0.953 (0.01) & 0.952 (0.01) & 0.943 (0.01) & 0.940 (0.01) & 0.946 (0.01) \\
 & MVP-VB-pop-pheno & 0.876 (0.01) & 0.910 (0.01) & 0.876 (0.01) & 0.898 (0.01) & 0.908 (0.01) & 0.917 (0.01) \\
 & MGflashfm & 0.772 (0.01) & 0.857 (0.01) & 0.654 (0.01) & 0.682 (0.01) & 0.692 (0.01) & 0.695 (0.01) \\
 & MultiSuSiE & 0.864 (0.01) & 0.860 (0.01) & 0.667 (0.01) & 0.671 (0.01) & 0.666 (0.01) & 0.669 (0.01) \\
 & mvSuSiE & 0.916 (0.01) & 0.914 (0.01) & 0.919 (0.01) & 0.914 (0.01) & 0.924 (0.01) & 0.930 (0.01) \\
 & MVLasso & 0.003 (0.00) & 0.003 (0.00) & 0.003 (0.00) & 0.003 (0.00) & 0.003 (0.00) & 0.003 (0.00) \\
\addlinespace[3pt]
\multirow{7}{*}{$15$} & MVP-VB & \textbf{0.910 (0.01)} & \textbf{0.934 (0.01)} & \textbf{0.904 (0.01)} & \textbf{0.894 (0.01)} & \textbf{0.919 (0.01)} & \textbf{0.931 (0.01)} \\
 & MVP-VB-pop & 0.901 (0.01) & \textbf{0.934 (0.01)} & 0.896 (0.01) & 0.892 (0.01) & 0.918 (0.01) & \textbf{0.931 (0.01)} \\
 & MVP-VB-pop-pheno & 0.865 (0.01) & 0.916 (0.01) & 0.850 (0.01) & 0.872 (0.01) & 0.895 (0.01) & 0.916 (0.01) \\
 & MGflashfm & 0.549 (0.01) & 0.663 (0.02) & 0.607 (0.01) & 0.643 (0.01) & 0.667 (0.01) & 0.691 (0.01) \\
 & MultiSuSiE & 0.772 (0.01) & 0.834 (0.01) & 0.642 (0.01) & 0.651 (0.01) & 0.673 (0.01) & 0.668 (0.01) \\
 & mvSuSiE & 0.706 (0.01) & 0.730 (0.01) & 0.717 (0.01) & 0.699 (0.01) & 0.726 (0.01) & 0.725 (0.01) \\
 & MVLasso & 0.005 (0.00) & 0.005 (0.00) & 0.005 (0.00) & 0.005 (0.00) & 0.005 (0.00) & 0.005 (0.00) \\
\midrule
\multicolumn{8}{l}{\textit{III. Shared causal SNPs with population-dependent trait associations.}} \\
\midrule
\multirow{7}{*}{$10$} & MVP-VB & \textbf{0.933 (0.01)} & \textbf{0.926 (0.01)} & \textbf{0.940 (0.01)} & \textbf{0.941 (0.01)} & \textbf{0.929 (0.01)} & \textbf{0.953 (0.01)} \\
 & MVP-VB-pop & \textbf{0.933 (0.01)} & 0.924 (0.01) & \textbf{0.940 (0.01)} & 0.939 (0.01) & \textbf{0.929 (0.01)} & 0.952 (0.01) \\
 & MVP-VB-pop-pheno & 0.872 (0.01) & 0.903 (0.01) & 0.844 (0.01) & 0.897 (0.01) & 0.897 (0.01) & 0.920 (0.01) \\
 & MGflashfm & 0.025 (0.01) & 0.036 (0.01) & 0.637 (0.01) & 0.606 (0.01) & 0.621 (0.01) & 0.623 (0.01) \\
 & MultiSuSiE & 0.628 (0.01) & 0.633 (0.01) & 0.650 (0.00) & 0.632 (0.01) & 0.637 (0.00) & 0.642 (0.00) \\
 & mvSuSiE & 0.897 (0.01) & 0.912 (0.01) & 0.922 (0.01) & 0.900 (0.01) & 0.919 (0.01) & 0.936 (0.01) \\
 & MVLasso & 0.003 (0.00) & 0.003 (0.00) & 0.003 (0.00) & 0.003 (0.00) & 0.003 (0.00) & 0.003 (0.00) \\
\addlinespace[3pt]
\multirow{7}{*}{$15$} & MVP-VB & \textbf{0.912 (0.01)} & \textbf{0.925 (0.01)} & \textbf{0.901 (0.01)} & \textbf{0.903 (0.01)} & \textbf{0.915 (0.01)} & \textbf{0.944 (0.01)} \\
 & MVP-VB-pop & \textbf{0.912 (0.01)} & 0.924 (0.01) & 0.893 (0.01) & 0.900 (0.01) & \textbf{0.915 (0.01)} & 0.943 (0.01) \\
 & MVP-VB-pop-pheno & 0.860 (0.01) & 0.908 (0.01) & 0.852 (0.01) & 0.869 (0.01) & 0.889 (0.01) & 0.919 (0.01) \\
 & MGflashfm & 0.006 (0.00) & 0.011 (0.01) & 0.529 (0.01) & 0.529 (0.01) & 0.544 (0.01) & 0.587 (0.01) \\
 & MultiSuSiE & 0.537 (0.01) & 0.661 (0.01) & 0.647 (0.00) & 0.637 (0.00) & 0.636 (0.01) & 0.643 (0.00) \\
 & mvSuSiE & 0.704 (0.01) & 0.730 (0.01) & 0.705 (0.01) & 0.702 (0.01) & 0.715 (0.01) & 0.733 (0.01) \\
 & MVLasso & 0.005 (0.00) & 0.005 (0.00) & 0.005 (0.00) & 0.005 (0.00) & 0.005 (0.00) & 0.005 (0.00) \\
\bottomrule
\end{tabular}
\end{table*}

In this setting, MVP-VB achieves the highest or tied-highest reported
mean $F_1$ in Scenarios~II and III for all populations. In Scenario~I, its performance is close
to that of MVP-VB-pop, which models the phenotypes jointly within
each population. The lower population-specific $F_1$ scores of MGflashfm and
MultiSuSiE in Scenario~I are consistent with the absence of
shared causal SNPs and the use of their common multi-group
selections to evaluate disjoint population-specific supports. For example, when $K=4$ and $S=15$, the mean $F_1$
scores for Population~1 in Scenario~II are $0.904$, $0.896$, and
$0.850$ for MVP-VB, MVP-VB-pop, and MVP-VB-pop-pheno, respectively.
The corresponding values in Scenario~III are $0.901$, $0.893$, and
$0.852$. Thus, in these examples, MVP-VB exhibits modest additional
gains over MVP-VB-pop and larger gains over MVP-VB-pop-pheno.

\begin{figure*}[t]
  \centering
  \includegraphics[
    width=0.96\textwidth,
    keepaspectratio
  ]{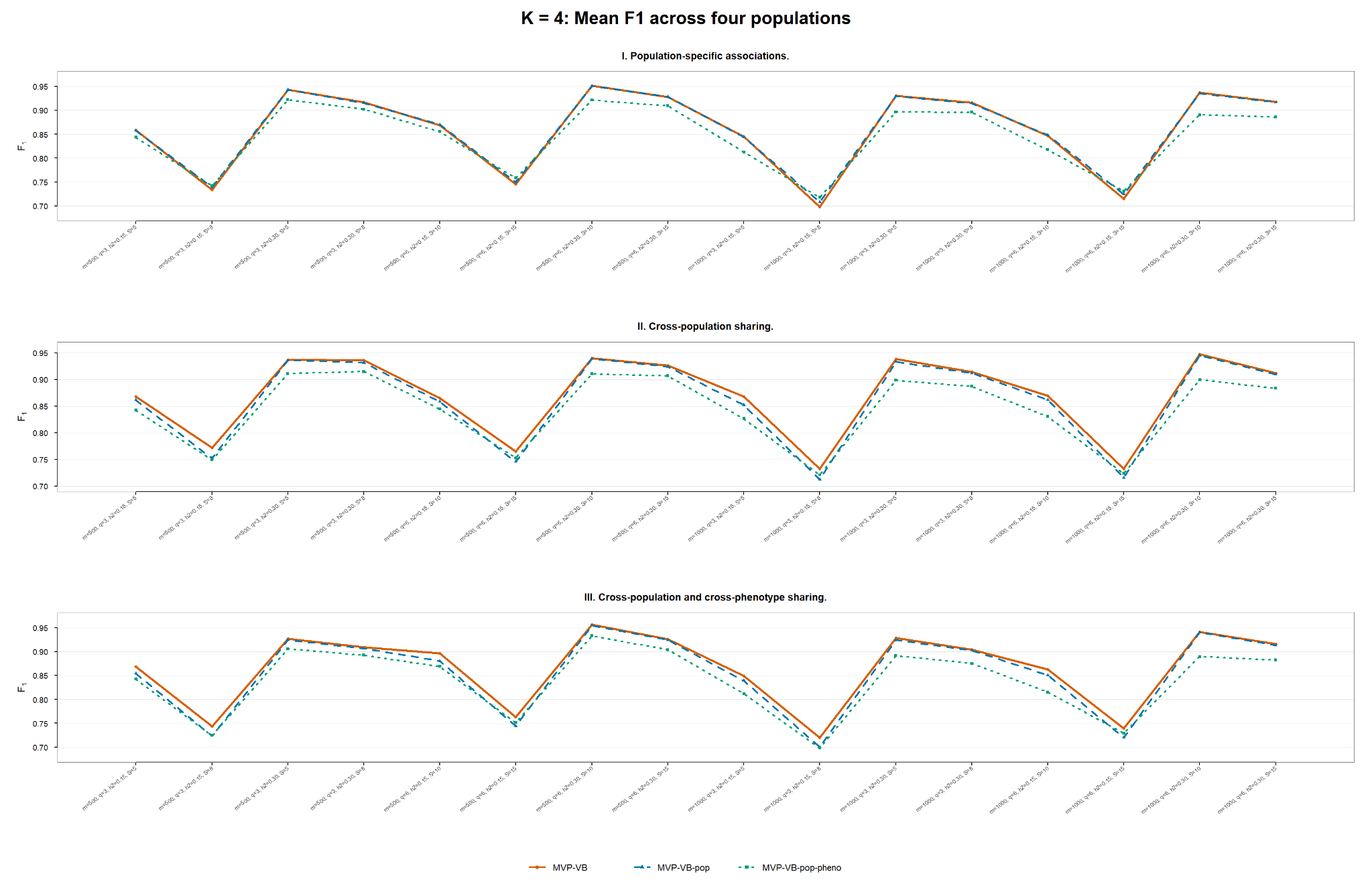}
  \caption{Mean $F_1$ scores of the three MVP-VB implementations for $K=4$.}
  \label{fig:k4-three-scenarios}
  \par\smallskip\noindent\textit{Alt text:} Comparison of mean $F_1$
  scores for MVP-VB, MVP-VB-pop, and MVP-VB-pop-pheno across
  three simulation scenarios and four-population configurations.
\end{figure*}

To examine this comparison across the simulation grid,
Figure~\ref{fig:k4-three-scenarios} presents the mean $F_1$
scores of the three MVP-VB implementations, averaged equally across
the four populations for $K=4$, under all three scenarios.
Each scenario contains 16 distinct $(m,q,h^2,S)$ combinations.
In Scenario~I, MVP-VB performs nearly identically to MVP-VB-pop,
indicating little loss in selection accuracy when cross-population
sharing is absent. In Scenarios~II and III, MVP-VB achieves the
highest four-population mean $F_1$ in all 16 configurations under
each scenario, demonstrating consistent advantages when
cross-population sharing is present. The improvements relative
to MVP-VB-pop are generally smaller than those relative to
MVP-VB-pop-pheno.

Furthermore, we report the runtimes of all compared methods in Table~\ref{tab:runtime}. 
\begin{table}[t]
\centering
\caption{Runtime (in seconds) of the compared methods. }
\label{tab:runtime}
\begin{tabular}{lcc}
\toprule
Method & \(K=2\) & \(K=4\) \\
\midrule
MVP-VB             & 6.30   & 30.70   \\
MVP-VB-pop         & 4.29   & 24.09   \\
MVP-VB-pop-pheno   & 3.68   & 13.96   \\
MGflashfm          & 250.82 & 1568.90 \\
MultiSuSiE         & 5.17   & 12.87   \\
mvSuSiE            & 7.84   & 25.01   \\
MVLasso            & 18.06  & 257.70  \\
\bottomrule
\end{tabular}
\end{table}

\section{Application to GLGC 2021}
\label{sec:glgc}

We applied MVP-VB to ancestry-specific GWAS summary statistics from the
2021 Global Lipids Genetics Consortium (GLGC) study \citep{Graham2021}
for four blood lipid traits: high-density lipoprotein cholesterol
(HDL-C), low-density lipoprotein cholesterol (LDL-C), total cholesterol
(TC), and triglycerides (TG), across five ancestry groups: African
(AFR), East Asian (EAS), European (EUR), Hispanic (HIS), and South
Asian (SAS). The final analysis set consisted of 190,356 SNPs, which
were retained after summary-statistic quality control, allele
harmonization, and matching to ancestry-specific LD reference panels
across all five ancestry groups and all four lipid traits.

The ancestry-specific sample sizes are
$n_{\mathrm{AFR}}=76{,}052$,
$n_{\mathrm{EAS}}=66{,}073$,
$n_{\mathrm{EUR}}=1{,}004{,}754$,
$n_{\mathrm{HIS}}=27{,}803$, and
$n_{\mathrm{SAS}}=32{,}138$.
Following standard LD-informed block-wise and overlapping-region
strategies used in GWAS analyses
\citep{BerisaPickrell2016,ChenEtAl2021DENTIST,FernandoEtAl2017},
the retained SNPs were analyzed in overlapping genomic regions to
reduce LD boundary effects.

For each ancestry group, the marginal-effect matrix
$\hat{\boldsymbol B}_k$ was constructed from A1-aligned GWAS effect
estimates after scaling by the expected genotype standard deviation.
The trait correlation matrix $\boldsymbol L_k$ was estimated from
LD-pruned genome-wide univariate GWAS $Z$-statistics following
\citet{LiuLin2018}, and $\boldsymbol D_k$ was approximated using
ancestry-matched 1000 Genomes Project Phase~3 HapMap3 LD reference
panels \citep{1000Genomes2015}. We then applied the summary-data
implementation described in Section~\ref{subsec:mvp-summary}.

We used Option~4 in Section~\ref{subsec:mvp-automatic}
for the joint five-population analysis.
Given the substantial differences in population sample sizes,
we applied the evidence-only marginal-statistic adjustment
described in the Supplementary Material when determining the
sharing structure, without changing the likelihood inputs.
Relative to the simulations, the initialization seed fraction,
effect-equivalence margin, and marginal-association threshold
for cross-population support were reduced to $0.001$, $0.01$,
and $0.01$, respectively.
For comparison, we also fitted MVP-VB to the European ancestry
group alone.

Using a posterior inclusion probability threshold of \(0.5\), the joint
five-population analysis selected \(327\) distinct SNPs, corresponding
to \(631\) population--SNP--trait combinations, whereas the EUR-only
analysis selected \(306\) SNPs and \(547\) SNP--trait pairs. Of the
\(306\) EUR-only SNPs, \(305\) were retained in the joint analysis.
The joint analysis additionally selected \(22\) SNPs, of which \(20\)
were selected exclusively in one or more non-European populations.

Using the historical lipid-association reference compiled by
\citet{Graham2021} from previously published variants and the GWAS
Catalog (accessed May 2020), we classified a locus as unmatched to the
historical reference when its lead SNP was more than \(500\) kb from
every reference variant. Under this criterion, the \(327\) selected SNPs
formed \(181\) loci, of which \(47\) were unmatched to the historical
reference. Among these \(47\) loci, \(31\) overlapped loci subsequently
classified as new in GLGC 2021, one overlapped a GLGC 2021 known locus,
and \(15\) were more than \(500\) kb from all \(941\) loci reported in
GLGC 2021. Among these \(15\) loci, one also showed a marginal
ancestry--trait association reaching \(P<5\times10^{-8}\). 

\section{Discussion}
\label{sec:conclusion}

In this article, we propose MVP-VB, a variational Bayesian method for
variable selection in multivariate regression across multiple populations.
Through alternative prior-sharing structures for the inclusion indicators,
the framework shares information across related regressions while
accommodating shared, population-specific, and response-specific
associations. Under appropriate conditions, we establish selection
consistency and coefficient estimation guarantees for fixed-dimensional
variational iterates, together with model-selection consistency and
coefficient contraction for the globally optimal variational posterior
in high dimensions. Numerical experiments demonstrate the benefits of
information sharing for variable selection, and the GWAS application
illustrates the use of MVP-VB with summary statistics.

Several directions merit further investigation. The current formulation uses a Gaussian likelihood. A natural extension is to develop multivariate logistic or probit models for correlated binary responses, together with corresponding variational algorithms and theoretical guarantees. Further work could examine the calibration of variational credible intervals and develop inference for differences in effects across populations. Another important direction is to develop theoretical guarantees for summary-data analyses that replace sample cross-products with external reference matrices, thereby quantifying the misspecification introduced by such substitutions.

\section*{Acknowledgment}
This study was funded by the National Natural Science Foundation of China (Grant No. 12401377).

\section*{Supplementary material}
The Supplementary Material provides derivations of the variational updates and the evidence lower bound, details of the summary-data implementation, and proofs of the fixed- and high-dimensional results. It also contains additional simulation results.
Code and analysis materials for MVP-VB are available at
\url{https://github.com/yyh0507/MVP-VB}.

\bibliographystyle{plainnat}
\bibliography{references}
\end{document}